# An Analysis of Chinese Search Engine Filtering

Tao Zhu Independent Researcher
zhutao777@gmail.com

Christopher Bronk Baker Institute for Public Policy Rice University
rcbronk@rice.edu

Dan S. Wallach Department of Computer Science Rice University
dwallach@cs.rice.edu

## ABSTRACT

The imposition of government mandates upon Internet search engine operation is a growing area of interest for both computer science and public policy. Users of these search engines often observe evidence of censorship, but the government policies that impose this censorship are not generally public. To better understand these policies, we conducted a set of experiments on major search engines employed by Internet users in China, issuing queries against a variety of different words: some neutral, some with names of important people, some political, and some pornographic. We conducted these queries, in Chinese, against Baidu, Google (including google.cn, before it was terminated), Yahoo!, and Bing. We found remarkably aggressive filtering of pornographic terms, in some cases causing non-pornographic terms which use common characters to also be filtered. We also found that names of prominent activists and organizers as well as top political and military leaders, were also filtered in whole or in part. In some cases, we found search terms which we believe to be "blacklisted". In these cases, the only results that appeared, for any of them, came from a short "whitelist" of sites owned or controlled directly by the Chinese government. By repeating observations over a long observation period, we also found that the keyword blocking policies of the Great Firewall of China vary over time. While our results don't offer any fundamental insight into how to defeat or work around Chinese internet censorship, they are still helpful to understand the structure of how censorship duties are shared between the Great Firewall and Chinese search engines.

## Categories and Subject Descriptors

K.5.2 [**Legal Aspects of Computing**]: Governmental Issues—*Censorship*; H.3.3 [**Information Storage and Retrieval**]: Information Search and Retrieval

## 1. INTRODUCTION

Latest in a line of electronic communications technologies whose history reaches back to the mid-19th century, the Internet is revolutionary in its capacity to permit many-to-many communications across global span at enormously low cost. It is also widely considered an open system or commons and many in the West opine that an electronic extension of the free speech and press endorsed in the United States and Western democracies may be applied to it. For countries where government oversees media activity, including in the digital domain, the free and unfiltered flow of ideas is considered undesirable; these governments take technical steps to block this undesired Internet content by policy and by technical means.

Our work is interested in quantifying the degree to which Internet search results are filtered in these countries. Internet search engines heavily influence how Internet users locate and view content [11]. High ranking in search results pages heavily influences which web pages are visited and also carries something of an imprimatur of importance. If a page ranks highly, it may be a best fit for the desired knowledge (i.e., it has high page rank [4, 24]), however, search ranking algorithms may not be faithful to the intention of the user's search [16], whether due to deliberate search engine manipulation by independent "optimizers" (e.g., some web sites' rankings are artificially inflated through the an array of false links from bogus web pages on other sites) or due to political pressure from the government for a web search vendor to tamper with their natural results.

For this reason, we approached the research question of how one country, China, exerts its influence upon companies providing Internet search services for the Chinese market, both from within and externally. We want to know how these search engines exercised censorship or filtering in providing search results to the Chinese people. In this paper, we will describe a variety of different experimental methods that we devised to better understand these search engines' behaviors.

**Legal frameworks and search censorship.** The choice as to what a search engine should or should not elide from its "natural" search results is grounded in law, custom and other mores. Questions of indecency, illegality and inappropriateness vary across cultures and international boundaries. Even among the legally-harmonized nations of the European Union, restrictions on content, for instance Germany's prohibition of neo-Nazi speech, are permitted. There is no universal maxim for free speech on the Internet, however, some countries are more permissive of broad unfiltered access than others.

As the early 2011 political upheaval in the Middle East indicates,

.

Internet connectivity can be a threat to regimes whose political systems balance on the controlled delivery of tailored propaganda by the political leadership. As a result, governments concerned with controlling the message have been increasingly drawn to technologies blocking unwelcome concepts, ideas or images from their citizens [39]. In many countries around the globe, control of information has extended from the traditional management of state-run news and media to newer active measures undertaken to remove or filter Internet information objectionable to the regime from public purview by technical means. Perhaps no country on the planet has taken greater action on this front than China. This is fodder for political debate and also serves as our core technological research challenge.

Whereas other nations have, at time of political trouble, chose to disconnect from the Internet, as Egypt did after the January 25 protests reached their peak, or in Burma during the abortive Saffron Revolution of 2007, China has rapidly grown its Internet user population without disconnecting from the world. As of 2010, the total number of Internet users in China reached 457 million, more than a third of the country, and representing an increase of 34.3% from the previous year [5]. The Ministry of Information Industry is responsible for the governance of China's rapidly growing Internet ecosystem in partnership with the Chinese Academy of Sciences via the China Internet Network Information Center, established in 1997.

In its period of exponential user base growth, rather than physically walling off its citizens from the Internet as other countries have done, China has attempted to temper the benefits of a generally open Internet with a variety of censorship tactics [20]. One group of researchers called it "a panopticon that encourages self-censorship through the perception that users are being watched" [28].China has invested heavily in a mix of Internet filtering technologies, network police surveillance, and restrictive regulations. All of these serve to tightly control access to undesirable Internet content.

From the beginning of widespread Internet adoption, the Chinese government has maintained a gap between its domestic internet space and the global internet [26]. Internet content filtering in China is clearly predicated in policy. Promulgated by the Ministry of Public Security in 1997, the China's Computer Information Network and Internet Security, in its Protection and Management Regulations, sets clear rules on the use of information. In addition, any citizen's use of computer networks is prohibited without prior approval. This policy provides the basis for the Golden Shield Project, China's national Internet firewall, colloquially called the "Great Firewall of China," and the recent Green Dam Youth Escort, reputedly an anti-pornography filter, required for installation on public-use computers in China.

Many countries use one or more of three censorship methods. The first of these is an infrastructure-dependent method to filter or block content and is made up of blacklists or other dynamic systems [6, 7, 25, 8, 35, 10]. The second is a user-focused approach whereby cyber police "maintain order in all online behaviors" [29] and functionaries called "Internet commentators" surreptitiously shape public opinion [36]. The third is company-oriented and is demonstrated through pressure to self-censor. Through strict regulations, the Chinese government imposes its will on Internet service providers, blogging sites, search engines, and others that stray from the dictated conventions. We have chosen to examine how that will may be impacting search engine operations.

**Contributions of this paper** This paper quantifies Chinese search engines' self-censorship, comparing search results from 8 different search engines, crawling more than 45,000 keywords over a period of 16 months. These data allow us to ask several interesting ques-tions.

- How does the Chinese government control or regulate domestic search engines?
- Do all search engines follow the same filtering policies?
- Is it possible that users know if the search engines they are using are practicing self-censorship?
- Are there easy workarounds for users to gain search results that have not been subject to the same degree of filtering?

## 2. CHINESE WRITING, IN BRIEF

To help non-Chinese readers of this paper, we now summarize several salient features of Chinese languages and how search engines must deal with their peculiarities.

About one-fifth of the world's population, or over one billion people, speak some form of Chinese as their native language. "Standard Chinese" is essentially the Mandarin Chinese dialect spoken natively in Beijing. Other cities speak very different dialects and two Chinese speakers from different cities may be completely unable to understand one another. Even with the broad diversity of spoken Chinese, written Chinese is essentially standardized. Most Chinese people who cannot speak to one another can still communicate in writing.

There are currently two systems for Chinese writing. The *traditional system* is used mostly in Chinese speaking communities outside mainland China. The traditional system descents from character forms dating back to the 5th century AD. Some traditional Chinese characters, or derivatives of them, are also found in Korean and Japanese writing. The *simplified system*, introduced in China in 1950s with the intent of promoting mass literacy, replaced most complex traditional glyphs with newer glyphs having fewer strokes. Simplified Chinese is also used in Singapore and Malaysia and is the most popular Chinese writing system, worldwide. For our work in this research, we focused entirely on simplified Chinese.

Chinese characters are derived from several hundred simple pictographs and ideographs in ways that are logical and easy to remember. *zigen* are the atomic components of Chinese characters. Some zigen represents meaning, while other zigen represents pronunciation. Unsurprisingly, Chinese characters sharing the same zigen usually have similar pronunciations or meanings.

Interestingly, Chinese people take advantage of this to defeat keyword-based censorship, replacing one character in a word with another sharing a similar shape. For example, "法轮功" (Falun Gong, a group which is broadly forbidden within China) is sometimes written as "法论功." Making seemingly minor changes to the second character of the word yields a result that's still perfectly legible to humans but can confuse automated censorship systems, at least until their human managers catch on.

Characters form the basic unit of meaning in Chinese, but not all characters can stand alone as a word; most Chinese words are formed of two or more characters. For example, the word "中华人民共和国" (People's Republic of China) is seven characters long and has smaller words within: "人民" (people) and "共和国" (republic country). The first two characters,"中华" are usually not be used as a word independently in modern Chinese, though it can be used as a word in ancient Chinese. Digging further, within word "人民" (people), "人" is a word (human), but "民" (civilian or folk) is not a standalone word.

English speakers expect that words are separated by whitespace or punctuation. In Chinese, however, words are simply concatenated together. Consequently, the problem of mechanically segmenting Chinese text into its constituent words is a difficult problem, and each search engine will necessarily employ different al-gorithms and heuristics toward this problem. As another example, while the proper segmentation of "中华人民共和国外交部" (Ministry of Foreign Affairs of the PRC) is "中华人民共和国 / 外交部", another word, "国外" (overseas), could also be erroneously extracted. Consequently, a search for "国外" should most likely not match the string "中华人民共和国外交部" but a query for "外交部" should.

Of course, sometimes search engines get this wrong. Search users are given something of an out by using quotation marks explicitly when they search. Quotation marks around a string direct the search engine to find the precise quoted characters consecutively, regardless of their surrounding context, thus bypassing the normal segmentation process. See §4.2 for discussion and experimental measurements on quotation.

## 3. EXPERIMENTAL SETUP

In order to gain a deeper understanding of the functionality and the extent, we investigated Chinese Internet search engine result filtering. Some amount of censorship is immediately obvious; some search engines will literally announce that they are withholding results. Likewise, we can observe obvious effects like TCP reset packets arriving which kill our session when we make specific queries.

We can also make differential comparisons across search engines, particularly when they are using the same underlying algorithms (e.g., Bing.com and cn.Bing.com can be expected to have similar or identical databases, as can Google.cn and Google.com). If one web site reports more hits than another for the same query, that's indicative of censorship.

Following this path of inquiry, we crawled roughly 45,000 different keywords on four well-known search engine companies operating in China and added new search websites: Baidu (Baidu.com), Google (Google.com, Google.cn and Google.hk), Microsoft (Bing.com and cn.Bing.com) and Yahoo! (Yahoo.com and cn.Yahoo.com)[1].

**Baidu** (百度, literally meaning "hundreds of times," also represents the persistent search for the ideal). Baidu is the most popular domestic search engine in China.

**Google** (谷歌, meaning "songs for millet or corn"). Before Google launched a Chinese local presence in February 2006, Google.com had been unavailable roughly 10% of the time [34]. In order to launch Google.cn, Google apparently agreed to remove "sensitive" information from their search results. Despite this, Google has never been the top search engine in China. They held roughly 30% of the market in 2009, which dropped to 26% in 2010, whereas Baidu's market share increased from 69% to 73% over the same time period [19].

**Yahoo!** (雅虎, meaning "elegant tiger"). Yahoo! China is not controlled by the U.S.-based Yahoo Inc. Instead Alibaba Group took control of Yahoo! China as part of a 2005 deal with Yahoo. Unlike Google or Microsoft, which keep confidential records of their users outside mainland China, Yahoo! China stated that the company does not protect the privacy and confidentiality of its Chinese customers from the authorities [15]. Yahoo! China was prominent in China search engine market share in the early 2000's, but it's market share decreased rapidly from more then 40 percent in 2003 to 0.3 percent in 2010 [18].

**Bing** (必应, meaning "must respond") Microsoft launched Bing

China's beta version in June 2009. Bing has less than 1% of Chinese market share in 2010.

To perform our experiments, we prepared *word sets* from which we form queries to searching engines. We build a crawler to visit the various search engines and make queries, and we build various analysis tools to extract the information we present later.

### 3.1 Word sets

All major search engines have rate limiting features, which requires us to be clever in how we design the set of queries we use for our experiments. If we use too many different words, then it will take too long between different instances of the same query. If we don't use enough words, we might miss something noteworthy.

As something of a control group, we need non-sensitive terms that are popularly used by Chinese Internet searchers. Zhongguo-osou.com conveniently collects the most popular keywords user searched in Baidu.com and Google.cn [37]. In total, there are 66, 516 words in this list, of which we use the first 44,102 words[2]. We will later use the term *General Words* to refer to these words.

We also add the word set which are known as sensitive words. Jedidiah et al.'s ConceptDoppler [8] discovered 133 specific words which are filtered in any HTTP GET request passing through the Great Firewall of China. We will later use the term *ConceptDoppler* to refer to these words.

We also included a variety of terms that might have been interesting in the future, in the hopes that we would be able to observe censorship in action. To that end, we used a list of 1,126 leaders within the Chinese government as well as the names of various Chinese government bodies and committees. We will later use the term *Leader Name* to refer to these words.

Additionally, we gradually added new words which we thought might be interesting to test, based on headline news from current world events as our experiments progressed, ultimately ending up with 85 such terms. We will later use the term *MyList* to refer to these words.

Some words occur more than once in the above sets. Once merged together, there were a total of 45,411 different words in our word set.

### 3.2 Crawler

Our crawler program is straightforward. The crawler takes a word from the word set, forms a query, sends it to a search engine, and saves the returned HTML file to local file system. We used the `wget` utility to simulate a web client, allowing us to automate the data collection process. We note that we generated a `user-agent` string from Firefox, working around some search engines that otherwise rejected our queries. Likewise, we had to properly manage the cookies set by some other search engines.

Our work initially began in early 2010, using words from General Words to probe for differences between Google.cn and Google.com. We only recorded the number of hits reported for each query and otherwise deleted the HTML files that came back to us. After March 22, 2010, when Google killed Google.cn, we decided to become more systematic in our efforts, adding 5 more search engines (Baidu.com, cn.Bing.com, Bing.com, cn.Yahoo.com, Yahoo.com) to our experiments, increasing the word sets by adding ConceptDoppler, MyList and Leader Name, and saving the full HTML responses we received on each query. We also conducted every query both with and without quotation marks around it (see §4.2).

---

To deal with search engine rate limiting, we needed to limit our query rate. Of course, we could use multiple IP addresses, simultaneously, since query throttling appears to be implemented on a per-IP-address basis, as far as we've seen it. Toward that end, we build a parallel crawler using 10 client PCs, each of which we allowed to have one outstanding query for each search engine. All of this was controlled from a central machine which doled out the query tasks. Results were written the query machines' local filesystems and later gathered together for analysis.

We set different querying intervals for different search engines. With the servers of Baidu.com and cn.Yahoo.com being physically located in China, the network latency and throughput are much lower than we observed for other search engines. We found that we did not need to introduce artificial delays. Instead, for each crawler machine we could maintain one query to each engine, non-stop. Despite this, it still took about 20 hours for a full test trail for Baidu.com and cn.Yahoo.com.

Yahoo.com uses the most strict robot detecting mechanism among these seven search engines. Even sleeping for 5 seconds after every query, our IPs were still blocked after about 30 minutes. Rather than stretching the sleep interval even farther, we instead sent queries sequentially, without delays, until we were blocked. Then we waited to be unblocked and resumed our crawling. Overall, this strategy required 11 days for a full crawl of our data set.

For Bing.com, cn.Bing.com, Google.com, and Google.hk, we settled on using a random sleep interval ranging from 0.7 to 2.2 seconds between each query. A full test trail for Google.com, Google.hk, Bing.com, or cn.Bing.com is about 10 hours. Bing.com and cn.Bing.com used to have weaker anti-crawling feature, but these were upgraded in November 2010. We originally could query our entire data set in an hour.

In the process of implementing our crawler, a variety of different things could go wrong that we needed to detect and manage. We group the errors into four classes:

**TCP Reset.** This appears to be caused by the Golden Shield Project, colloquially called the Great Firewall of China (GFC), and is triggered by querying sensitive words. With wget, these TCP reset packets usually manifest themselves with "Read Error (connection reset by peer)" or "Read Error (Connection timed out)". Typically, the GFC then blocks all communication from our IP address to the search engine for roughly 90 seconds. When we get in this state, we pause ten seconds then query once every ten subsequent seconds with a "Hello world" query until we either get through or get a different error. TSP reset errors happen mostly to cn.Yahoo.com, but they also happen to Baidu.com with a much lower frequency. (More details in §4.4.)

**HTTP error.** When Yahoo detects us as a robot and wishes to throttle us, it sends back "ERROR 999: Unable to process request at this time". We stop for 5 minutes before testing again with a "Hello, world" query. When Google detects us as a robot, it sends back "503 Service Unavailable". As before, we fall back and retry with a "Hello, world" query until we get through.

**HTML error.** For Baidu and Bing, when they detect us as a robot, they do not return an HTTP error code. Instead, they return an HTML page with suitably apologetic text. These HTML pages changed on a regular basis during our experiments, requiring us to make suitable modifications to the crawler.

**Timeout.** These happened for reasons that we cannot diagnose. We treated timeouts as temporary errors, waited ten seconds, then attempted a "Hello, world" query to ensure that everything was working again.

If any of these error conditions occur, we fall back to making sure out "Hello, world" query succeeds and then we retry the query that induced the error. We try up to 3 times before we give up on a keyword.

Naturally, there were many unpredictable difficulties during the experiments. For example, different search engines use different encoding system for Chinese characters, and the encoding systems are not same with the return pages and the queries. cn.Yahoo.com return the results in UTF-8, however the query has to be encoded in GB18030. Similarly, Baidu.com returns the results in the form of GB2312, however, we have to send our query in GB18030. cn.Bing.com, Google.hk and Google.cn use UTF-8 in both results and queries.

## 3.3  Analyzer

Our analyzer is designed to parse the HTML files returned by our crawler, making use of Beautiful Soup, a Python library for parsing XML-style documents. In sum, we collected one terrabyte of HTML files (uncompressed). We parallelized our analyzer, running on 16 CPUs, and the full computation took roughly 100 hours to run, dumping the resulting data into a MySQL database which we can more easily process.

In the process of debugging our analyzer, we had to deal with the ever-changing layout of the various search engines' pages as well as a variety of transient error conditions. Many errors only became apparent when trying to understand strange artifacts in our graphs.

## 4.  DETECTION METHODS

We now describe several different measurement experiments and our findings. Table 1 summarizes how many different measurements we made of each search engine, where one measurement corresponds to queries made to that engine for each word in our corpus. Numbers in parentheses count how many measurements we made before Google terminated Google.cn.

### 4.1  Hit ratios

When a user queries a search engine, the search engine typically says how many results match the query (see Figure 1). This is true for every major search engine. Our hypothesis is that this number may be useful as a way of measuring search engine censorship.

Of course, the number of hits for a given query is not meaningful in and of itself. However, the same query sent to two different search engines will allow us to measure a *hit ratio*. In a world without any censorship, we would expect this ratio to be roughly 1.0, regardless of query, assuming the two search engines use the same underlying database or underlying codebase. Of course, there will be noise in these measurements. We have observed a 10% variation in these results in repeated queries for the same term, even for identical queries made within minutes of one another to the same search engine. Likewise, we can imagine that there will be some variation in search results that are a function purely of the way the search engine processes Chinese-language queries.

Censorship could manifest itself on the input to a search engine via its crawler, perhaps a result of the crawler being forced to view the Internet through the Great Firewall of China. Censorship could also manifest itself on the output of a search engine via internal policing. If a crawler were censored as it gathered its data, then the number of results reported would necessarily be lower. If censorship was implemented internally, a search engine could perhaps present the full number of results yet quietly fail to return matching results.

**Table 1: Experimental measurement runs: how many times each corpus of words were queried against each search engine.**

| | Words list | Number of words | Google .com | Google .cn | Google .hk | Baidu .com | Bing .com | Cn.Bing .com | Yahoo .com | Cn.Yahoo .com |
|---|---|---|---|---|---|---|---|---|---|---|
| No quotation marks | General Words | 44102 | 19 (4) | (5) | 18 | 3 | 6 | 6 | 2 | 2 |
| | Leader Name | 1126 | 21 | - | 20 | 3 | 6 | 6 | 2 | 2 |
| | ConceptDoppler | 133 | 21 (2) | (3) | 20 | 3 | 6 | 6 | 2 | 2 |
| | MyList | 85 | 7 | - | 6 | 3 | 4 | 4 | 2 | 2 |
| With quotation marks | General Words | 44102 | 19 (3) | (3) | 18 | 17 | 23 | 23 | 12 | 15 |
| | Leader Name | 1126 | 21 | - | 20 | 18 | 23 | 23 | 14 | 17 |
| | ConceptDoppler | 133 | 21 (3) | (3) | 20 | 18 | 23 | 23 | 15 | 16 |
| | MyList | 85 | 7 | - | 7 | 6 | 7 | 7 | 5 | 5 |

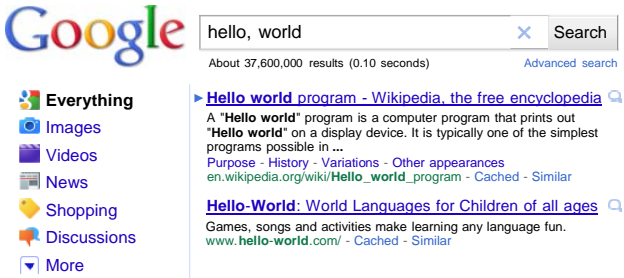

**Figure 1: All search engines report an approximate number of pages matching a given query. Here, Google reports just over 37 million occurrences of the phrase "hello, world" on the Internet.**

We have also observed differences, for example, in the way that English and Chinese-language versions of the same search engine segment Chinese words for queries. Furthermore, the English-language search engines, given Chinese characters for a query, will sometimes match Japanese-language pages, since some Chinese characters also show up in Japanese writing. This doesn't happen with the Chinese-language search engines.

Consequently, we expect that hit ratios will be a noisy signal, with a variety of reasons other than censorship which might induce ratios that are notably lower or greater than one. Regardless, as we will now show, hit ratios provide a valuable window into the world of search engine censorship.

### 4.1.1 Google

We performed five sets of measurements from August 2009 to March 2010, when Google ultimately shut down Google.cn. Figure 2 shows cn/com ratios for querying our corpus against Google, without use of quotation marks. Each dot in the picture is a specific search term. Five different colors are used for the five different measurement sets. We sorted the results based on the cn/com ratio (lowest to highest) and the x-axis position is the location in this list. The y-axis position, in log-scale, indicates the cn/com ratio. Since the rank ordering of the ratios would be different on each measurement run, we ordered all the results based on the median cn/com ratio. Thus, each of the five points in a given column represent five queries for the same word at different times. Also, we plotted the median cn/com ratio.

This plot indicates something of a fuzzy band between the cn/com ratios of 0.1 and 10.0 for the vast majority of queries. We see a similar effect in other measurements. This seems to indicate that *ratio differences within a factor of ten in either direction are not a significant indicator of tampered results*. Instead, these are the results of the many other factors (see §4.1 for some possibilities).

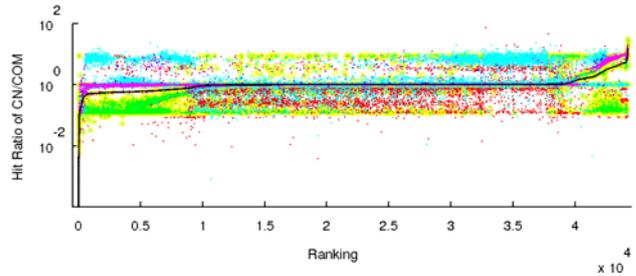

**Figure 2: cn/com hit ratio for Google (no quotation) over five different measurements of our word set from August 2009 through March 2010.**

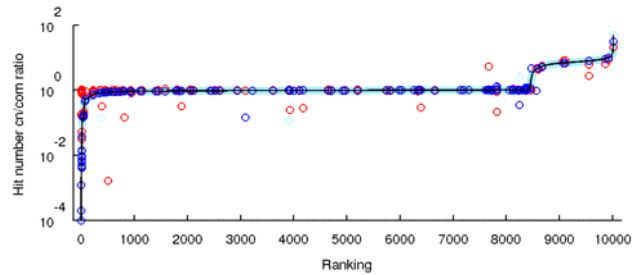

**Figure 3: cn/com hit ratio for Google (with quotation) over three different measurements of our word set from January 2010 through March 2010.**

Despite this, there are clear "tails" on both sides of the graph. We're particularly interested in the search queries with the lowest ratios; the lowest is for "柴玲" (Chai Ling, one of the leaders of the Tiananmen Square protests in 1989). Google.cn reports a grand total of 34 hits for her, while Google.com reports 230,000 hits (a ratio of 0.00014). Unsurprisingly, Chai Ling is widely censored in China. When we examined the other searches with a cn/com ratio lower than 0.1, they were generally either political or pornographic in nature. The "boundary" where censored terms start running into the noise is roughly around word #66 in the list, at which the ratio is roughly 0.12. We present these numbers, and those for other search engines, in Table 2, columns A through D.

We also conducted three sets of measurements against Google.cn and Google.com using quotation marks around our search terms, as an experiment to see whether there was a meaningful difference between that and unquoted searches. We used a subset of 10,000 words from the General Words set. The cn/com ratio is plotted in Figure 3. This data has a similar shape to the non-quoted search terms, so we present it here. One notable difference between quoted and non-quoted searches for the same terms is that we did not see any differences in searches for pornographic terms. (Does this mean that Chinese users of Google.cn could get around porn censorship by using quotation marks around their searches? We, unfortunately, can't go back and try more variations on this to see what was going on.) Issues surrounding data quotation are discussed further in §4.2.

We also decided to look at the search terms on the other side of the graph, where Google.cn reported more search results than Google.com. There was no discernable pattern to these words. As far as we can tell, these terms' higher Google.cn hit rates was the result of minor or inconsequential differences between Google.cn and Google.com.

**Google hk/com.** When Google terminated Google.cn in March 2010, they relocated the simplified Chinese service to Google.hk alongside the Traditional Chinese service which had already been there. Google claim that, unlike Google.cn, Google.hk would no longer be subject to censorship. We began measuring the simplified Chinese version of Google.hk following the same methodology we used with Google.cn.

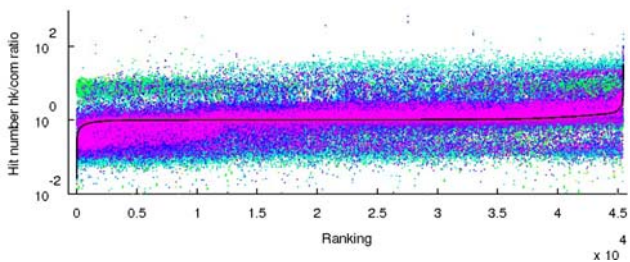

**Figure 4: hk/com ratio for Google (no quotation) over 25 data sets from May 2010 to February 2011**

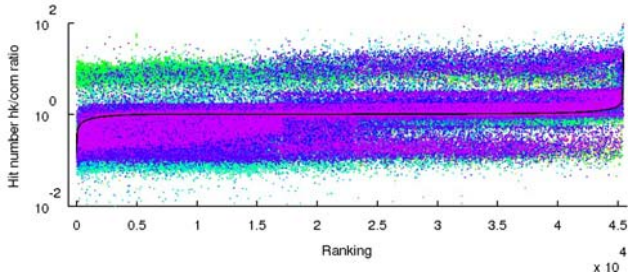

**Figure 5: hk/com ratio for Google (with quotation) over 25 data sets from May 2010 to February 2011**

The hk/com ratio and the median over all 25 measurements are plotted in graphs Figure 4 and Figure 5. Unlike the Google.cn data, the tail on the left side of the graph, minimizing the hk/com ratio, contains no particularly interesting words, with or without the use of quotation marks. Our results are thus consistent with an absence of externally-imposed censorship.

We note that Google.hk has no option to disable Google's *Safe Search* which is always "strict". When we used pornographic search terms with Google.hk, it returns a page explicitly saying "The word has been filtered from the search because Google SafeSearch is active." Also, no hit numbers are available for those search terms. For Google.com we left the default setting, which is "moderate", for our comparisons with Google.hk.

The right-side tails, when hk/com is maximized, turn out to be mostly pornographic words for both sets of measurements. We be-lieve this difference is due to the different ways that the different search engines treat pornographic terms. Google.hk, apparently, will take a term like "强奸犯" (rapist) and remove the first two characters (强奸, rape) doing a query on the last character, alone (犯, convict), thus yielding a larger number of results. In contrast, Google.com strictly searches on the more limited three-character term.

### 4.1.2 Yahoo!

For our analysis of Yahoo!, we only present results captured prior to August 2010, when yahoo.com abandoned its internal search engine in favor of Microsoft's Bing.

Figure 6 is the plot of cn/com ratio for Yahoo! before August 2010. Here we only show the results for queries using quotation marks. The distribution is clearly different from what we observed with Google. There are two relatively flat areas in the chart. One is around ratio of 0.1, another is between 10 to 100, with a notable discontinuity around the middle of our data set. In short, there are some terms where cn.Yahoo.com clearly dominates Yahoo.com and there are other terms where Yahoo.com clearly dominates cn.Yahoo.com. This leads us to believe that these search engines don't share much in the way of common infrastructure.

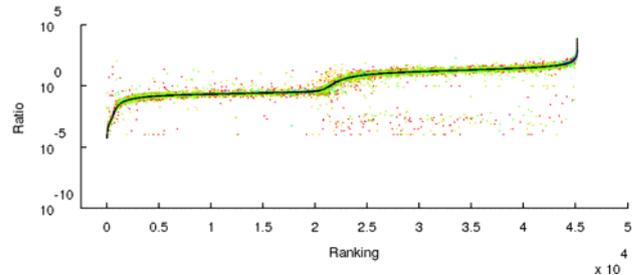

**Figure 6: cn/com ratio for Yahoo! (with quotation) over 8 data sets from May to August 2010.**

Many of the queries with the lowest cn/com ratio turn out to be pornographic terms. Interestingly, we observed that the character "" (meaning color or pornographic) is a particular focus of censorship attention from cn.Yahoo.com. Of the 120 words with that character in our corpus, 92 words have ratios lower than 0.1, 22 words between 0.1 to 0.3, and 6 words have rations of 0.3 to 0.86. The bulk of these words are not pornographic terms in any fashion at all. We hypothesize that the engineers of cn.Yahoo.com took on the relatively simplistic approach of filtering the all the keywords which contain this character, regardless of any collateral damage. To comfirm this theory, we looked for other characters in our that tend to show up in pornographic words: "插" (insert), "女" (female), and "射" (shoot/ejaculate). These characters, regardless of the actual word using them, all appear to induce cn.Yahoo.com to fall back to a relatively small set of web sites for its results. This leads us to hypothesize that cn.Yahoo.com implements a fallback strategy: any query that might possibly have sensitive results will only return web sites from a known-safe white list. We discuss this further in §5.1.

### 4.1.3 Bing

The cn/com method when applied to Bing (comparing Bing.com with cn.Bing.com) yields a plot that's quite similar to our comparison of Google.com and Google.hk. (To save space, we don't show it here.) This leads us to believe, unsurprisingly, that Bing.com and cn.Bing.com use the same underlying search engine. Examin-ing the words with cn/com ratio divergent from 1.0, at both ends, we could not find any particular pattern. But as we will discuss in §4.3, cn.Bing.com shows "removed results" banner message for many words, which explicitly tells users that censorship has been applied. Since the hit numbers are unchanged but censorship is clearly present on cn.Bing.com, either the number of censored pages is too small to measure, or the reported hit numbers on cn.Bing.com are simply inaccurate.

## 4.2 Quotation comparison

As we have discussed before, Chinese search engines, much like their English counterparts, behave differently when a query is put inside quotation marks. For example, Google's help page offers that, "by putting double quotes around a set of words, you are telling Google to consider the exact words in that exact order without any change."

For Chinese search engines, double quotes have a similar function, telling the search engine to consider the exact characters in the exact order. This is true for Google.hk and cn.Bing.com, But not always true for Baidu.com and cn.Yahoo.com. For example if you search for "共和国中华人民" (a reordering of the characters for "People's Republic of China"), in quotation marks, Baidu.com and cn.Yahoo.com will still return results for "中华人民共和国" (the proper spelling), but with fewer total results than you might get for the proper spelling.

Since the treatment of quotation marks around a strong clearly changes the results, we thought we should try quotation marks against our corpus to see whether sensitive words are treated differently when quoted. With this, we can both compare a Chinese and non-Chinese search engine (e.g., Google.cn vs. Google.com, as before) and we can also compare a search engine against itself.

### 4.2.1 Google.cn

Figure 7 shows the ratio of hit number of non-quotes vs. quotes for Google.cn. Though we only have two measurement sets for Google.cn, this graph clearly demonstrates that most searches yield comparable results (a ratio near 1.0). Only 188 words out of our whole set had a ratio < 0.9.

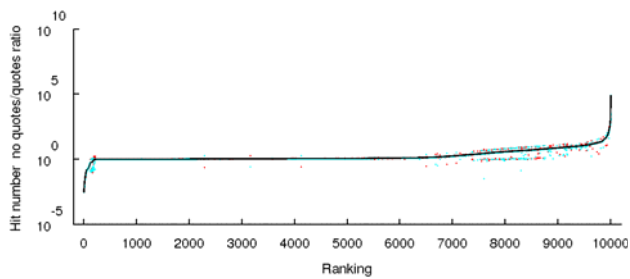

**Figure 7: Non-quotation vs. quotation hit ratio for Google.cn over two data sets in March 2010.**

The top 50 and next 100 low-ratio words from Google.cn are categorized in Table 2 in column E and F. The numbers in parentheses are the number of words with 4 or more characters. Our measurements show that sensitive words, mostly pornographic, were filtered from regular Google.cn results but appeared when they were in quotation marks.

In contrast, for Google.com and Google.hk, the top words were not especially sensitive. What did jump out were Chinese four-character idioms (typically bits of folk wisdom, or common sayings; English has many comparable sayings, e.g., "The early bird gets the worm."). Common Chinese idioms, when quoted, yield more search results than when they're not quoted. 61% of the top 100 ratios in Google.com and 69% of the top 100 ratios in Google.hk correspond to the prevalence of these Chinese idioms. Querying for common English sayings on Google.com and Google.hk, we observed similar results. When they are quoted, Google.com and Google.hk yield more search results than when they're not quoted. We won't attempt to understand why this effect is so significant, but it's clearly a common feature of Google search engines which is not specifically designed for the Chinese language and appears to be unrelated to any form of censorship.

### 4.2.2 Baidu.com

For Baidu, there are 83 words whose non-quotation/quotations hit ratio is lower than 1.0. We categorize these words in Table 2 in column G. We saw a split among pornographic terms, Tiananmen Square terms, and various important Chinese leaders and their relatives. In total, 82% of these specific words are sensitive. For the 83 words in question, Baidu's censorship filters clearly behaved differently. We discuss these terms further in §4.3, when we discuss search engines that explicitly disclose they've removed censored results from any given query result.

## 4.3 Removed Results Message

Every search engine we investigated will sometimes show a message which states that there are messages removed from the results when a search query would otherwise lead to a censored page. cn.Yahoo.com notably stands out by *always* saying "According to relevant laws, regulations and policies, some search results may not appear." (根据有 关法律法规和政策，部分搜索结果可能未 予显示。) regardless of the search query. Consequently, cn.Yahoo.com gives us no useful information about when it does and doesn't censor its results, but we learn something from the presence or absence of such a message on other search engines.

For Google.com, Google.hk and Bing.com, we found that none of the queries in our corpus would trigger such messages. For Google.cn, however, it was easy to trigger this information. Even querying for something innocuous like "你好" (hello), Google.cn informed us that one of the search results had been removed from the 10th page of search results (at 10 items per page). From this, we can conclude that Google.cn's index tagged all web pages with whether or not they're sensitive, rather than removing them entirely from the index.

### 4.3.1 Baidu

Unlike Google.cn, Baidu.com displays its "removed results" banner message as the first line of the first page of search results, regardless of how deep the censored search results may be. Furthermore, there is never such a message on the second or subsequent search results pages.

In our measurements with quotation marks, we found 35 words leading to a visible indication of censorship. In our measurements without quotation marks, we found 120 words leading to the same indication. This is consistent with our earlier results (see §4.2), where quoting search terms results in larger result sets. This seems to indicate that Baidu.com has two passes of censorship: one when the results to a query are first generated, and a second when the user tries to actually dig into the search results and see them. Table 2, in columns J and K, breaks down Baidu.com's censorship by type of query.

### 4.3.2 Bing

cn.Bing.com appears to take a similar censorship strategy to Goo-

gle.cn, as described above. Our experiments fetched the first 50 top hits for every word for each query, so our quantitative results are based on whether censorship appears in these top 50 hits. We would not observe censorship of lower-ranked search results.

First, for a subset of the words in our queries, we saw the "removed results" string every time, while for other words we only sometimes saw the "removed results" string. Table 2, columns H and I, reports statistics on the words that *always* have some censorship reported.

Also, we observed that roughly half (20,000 out of 45,000) of our queries gave search results which triggered page removal messages *at least once* among all of the trials for each given query. These correspond to approximate 38%, 32%, 88% and 74%, of the queries from the General Words, Leader Name, ConceptDoppler and MyList word sets, respectively. Unsurprisingly, word sets with more sensitive terms (ConceptDoppler and MyList) have a higher incidents of removed search results.

## 4.4 TCP reset

It's a standard practice for the GFC to snoop on TCP traffic, looking for undesired keywords. Its standard practice, when it sees such things, is to forge TCP reset packets to both sides of the connection, but it doesn't happen consistently.

One keyword that does result in a consistent, nearly 100% certain reset rate is the term "falun" in the standard Latin alphabet, not the equivalent Chinese characters. This happens regardless of the web site inside China that we might try to connect to, even government web sites. Perhaps they're more concerned with the term "falun" inbound to their country than they are with where, exactly, it's going.

Not all sensitive words have such a high reset rate as "falun". There are two search engines, inside the GFC that we can query from the outside, and their TCP reset behavior is different.

### 4.4.1 Baidu

We found 6 other words which consistently caused TCP resets when we sent them as queries to Baidu.com:

章沁生 Zhang QinSheng, recently promoted to a full general of the army.

彭小枫 Peng XiaoFeng, a full general of the army.

天葬 Sky burial, a Tibetan funerary practice.

我的奋斗 *Mein Kampf*, by Adolf Hitler, a book that's banned inside China.

王斌余 Wang Binyu, a Chinese migrant laborer executed for murder.

盛雪 Sheng Xue, a Chinese pro-democracy activist, journalist, and actress, currently living in Canada.

It is unknown to us why GFC resets these specific six words on Baidu. (These data are also summarized in Table 2, column L.)

### 4.4.2 Yahoo!

Among the words we tested, we found 102 words which caused TCP resets at least twice when sending them as queries to cn.Yahoo.com. Almost all of them are political sensitive words including terms related to the Tiananmen Square protests, Falun Gong, dissidents' names, and various Chinese dissident / pro-democracy organizations. Only three pornographic or obscene terms caused TCP resets. (These data are also summarized in Table 2, column M.)

Why is cn.Yahoo.com treated differently than Baidu.com? We conjecture that Baidu has implemented stronger internal censorship controls, or at least has convinced its government regulators that it has, and therefore Baidu has less overt restrictions placed on it by the GFC.

## 5. OTHER ANALYSES

In addition to our automated queries, we conducted a handful of manual experiments that helped us develop theories to explain how Chinese search engines implement their censorship behavior.

## 5.1 Sensitivity of different terms

Our search engine studies found clear evidence of censorship from search engines inside China (Google.cn, Baidu.com, cn.Yahoo.com) as well confirming other studies of GFC censorship. In this next section, we consider several specific sensitive terms and what this might say for the structure that the Chinese government may be requiring for internal search engines.

We queried "法轮功" (Falun Gong) at Google.cn and Google.com on March 20, 2010. Google.cn reported 16,900 hits versus Google.com's 880,000 hits (a 2% ratio). We then visited each web page returned by Google.cn in its first 20 results. Every page could be characterized as negative to Falun Gong. Furthermore, the top 100 hit results were from less than 20 unique top-level domains, each of which appears to be directly controlled by the Chinese government (e.g., people.com.cn, cctv.com, xinhuanet.com and china.com.cn). This suggests a "white list" policy for sensitive terms. When a search engine is given a sensitive query, it may only be permitted to return officially sanctioned results.

We also believe there to be a "second-class white list" of web sites, which have lower priority than those in "white list" but higher priority than other general web sites. For example when we search for pornography related terms on Google.cn, the resulting web sites are not limited to the "white list" sites we see elsewhere, but are still limited to a variety of govenment controled web sites.

We also believe there to be a "black list". These web sites may not be shown in search results under any circumstances. This was easy to validate. We visited epochtimes.com, the official homepage for Falun Gong, which is certainly censored within China. We extracted a specific sentence which we then fed to various search engines. Google.com found this sentence precisely three times, while Google.cn found it twice. The two common records were the same. The absent record was from epochtimes.com.

We observed the same white list phenomena in cn.Yahoo.com. 80 of the top-level domains of the first 100 returned results (fetched in May 2010) for the keyword "颜色" ("color" in English, and misjudged by cn.Yahoo.com to be a sensitive keyword, as we discussed in §4.1.2), are from exactly two top-level domains: ce.cn and china.com.cn. The other 20 hits include gb.cri.cn, chinadaily.com.cn, people.com.cn, xinhuanet.com and cctv.com, all of which belong to the Chinese government. For "屁眼" ("asshole" in English), measured in February 2011, cn.Yahoo.com returned 184 hits. Again, all of the top-level domains for the first 100 search results were from the above sites. We can safely conclude that these are all white-listed for cn.Yahoo.com. As an aside, we also observed that, some time in late 2010, cn.Yahoo.com improved their system such that many incorrectly sensitive words, such as "颜色" (color), are no longer treated as sensitive. Regardless, even today, we observe that cn.Yahoo.com still serves up results from the same white list whenever queried with a sensitive word.

Baidu.com also appears to have a broader white list than cn.Yahoo.com.

We finally consider whether Baidu.com, cn.Bing.com, and cn.Yahoo.com have comparable black list behavior to what we saw with Google.cn. We queried the same sentence from epochtimes.com as we originally tried with Google.cn. Baidu.com returned exactly one result, and it was from a non-governmental gamers' web site in Taiwan. There were only two hit results in cn.Yahoo.com, both from Buddhism forum within China. The article there containing the quoted sentence and was critical of Falun Gong. We could not observe the presence of black list in cn.Bing.com, which did link us back to the original source of the sentence on epochtimes.com.

**Table 2: Classification of the most censored words for different types of search engine analyses.**

| | cn/com ratio | | | | Quotation | | | Removed Result Message | | | | Reset | |
|---|---|---|---|---|---|---|---|---|---|---|---|---|---|
| | Google | | | | Google.cn | | Baidu | cn.Bing.com | | Baidu | | Baidu | cn.Yahoo |
| | no quote | | quoting | | | | | quoting | no quote | quoting | no quote | | |
| Word range | 1-66 | 67-100 | 1-15 | 16-100 | 1-50 | 51-100 | 1-83 | 1-53 | 1-59 | 1-35 | 1-120 | 1-6 | 1-102 |
| Ratio | <0.12 | <0.137 | <0.042 | <0.546 | <0.145 | <0.33 | < 1 | - | - | - | - | - | - |
| Pornography | 34 | 5 | 0 | 2 | 30(1) | 15 | 28 | 1 | 7 | 2 | 47 | 0 | 3 |
| Tiananmen Sq. | 14 | 0 | 10 | 4 | 4 | 1 | 14 | 18 | 11 | 11 | 22 | 0 | 35 |
| Falun Gong | 5 | 0 | 3 | 1 | 1 | 0 | 2 | 8 | 8 | 3 | 5 | 0 | 11 |
| Leaders | 2 | 1 | 0 | 4 | 1 | 6 | 21 | 3 | 2 | 17 | 39 | 2 | 6 |
| Politics | 7 | 1 | 2 | 4 | 2 | 1 | 3 | 11 | 6 | 1 | 6 | 4 | 44 |
| - | 4 | 27 | 0 | 70 | 12(6) | 28(4) | 15 | 12 | 25 | 2 | 1 | 0 | 3 |
| Total | 66 | 34 | 15 | 85 | 50 | 50 | 83 | 53 | 59 | 36 | 120 | 6 | 102 |
| Percent[1] | 94% | 21% | 100% | 18% | 76% | 44% | 82% | 77% | 58% | 94% | 99% | 100% | 97% |
| Column Name | A | B | C | D | E | F | G | H | I | J | K | L | M |

[1] Percent = Number of words in the five sensitive categories (Pornography, Tiananmen Sq., Falun Gong, Leaders and Poltics) / Total ×100%.

## 5.2 Temporal variation

We are interested in how search engine responses to sensitive queries change over time as well as how the behavior of the GFC changes over time. Clearly, comparing the number of hits for a given keyword is unlikely to be of much use, both due to the high variance or noise in hit counts, and due to the fact that search engines index new pages and update themselves on an ongoing basis. On the other hand, the TCP reset behavior of the GFC would likely be relatively stable over time, or at least between points when the administration makes adjustments.

**TCP resets.** In cn.Yahoo.com, during our experiments, we saw 184 different words trigger at least one TCP reset from the GFC. As we discussed earlier, there was some unpredictability in the GFC's behavior, but there were no clear trends visible in our data with one exception. For the measurements we performed on September 15, 2010, the GFC was notably less restrictive than on any other day in our data collection. We have no particular theory to explain why the GFC might have behaved differently on this one day.

Omitting the anomalous September 15, 2010 measurement, the only word whose reset rate was particularly notable for having a variable reset rate was "滕文生" (Teng Wensheng), a Chinese Community Party "think tank" researcher. Among our measurements from June 8, 2010 through February 21, 2011, his name triggered reset response continually in 4 sampling periods in August and September and did not trigger reset response before and after this period. This only happened when connecting to cn.Yahoo.com, not Baidu. We have no particular theory to explain this anomaly. (Maybe he's more important than we might have otherwise thought.)

**Removed results.** §4.3 demonstrated that this is an efficient method to detect sensitive words. Can this method also tell us some information about time trends?

Baidu.com notably gives us one bit of information (censored results within or not) as part of the initial result from a search query. This makes it easier to track trends over time than with other search engines where we would need to dig deeper in the search results to identify the presence of censored results. Among our full dataset of Baidu queries, there were 35 words that had censored results at least once. Among these 35 words, 26 words had censored results every time. The remained 9 words, described in Table 3, may be useful for measuring changes in censorship over time.

**Table 3: Trigger Removed Result Message Ratio in Baidu.com**

| Word (in English) | Censorship Observed | Total Observations | Ratio |
|---|---|---|---|
| Li Peng | 16 | 18 | 0.89 |
| Wen Jiabao | 12 | 17 | 0.71 |
| Zhang Dejiang | 6 | 18 | 0.33 |
| Chai Ling | 6 | 18 | 0.33 |
| Nobel Peace Prize | 1 | 3 | 0.33 |
| Month | 5 | 17 | 0.29 |
| Year Month | 4 | 17 | 0.24 |
| Zhou Xiaochuan | 4 | 18 | 0.22 |
| Adult | 2 | 17 | 0.12 |

Let's take "温家宝" (Wen Jiabao, the current prime minister of China) as an example. In our first five measurements (in June 2010), queries of "温家宝" did not trigger any notice of censorship, which means that, during this period, 温家宝 was not in Baidu.com's censorship list. However, in the 12 subsequent measurements (from July 2010 through February 2011), "温家宝" queries always indicated the presence of censorship. While we do not know what pages related to the prime minister might be subject to censorship, we do have a stable signal that censorship is being applied. We might then dig deeper to sort out what, exactly, has changed.

## 6. RELATED WORK

External evaluation of search engines does not require privileged access to any search engine's database or on specific knowledge of how the search engines work, and consequently many have studied them. In their first paper using external evaluation of search engines Bharat and Broder [3] picked pages uniformly at random from the index of a particular engine. They measured relative sizes and the overlaps of search engine indices through random queries. Bar-Yossef and Gurevich [1] published a more detailed study along similar lines, focusing on decreasing the biases during sampling. In later work [2] , they measured global quality metrics of search engines, like corpus size, index freshness and density of duplicates in the corpus. Vaughan [32] also evaluated search engine ranking quality and stability.

Our work is specifically focused on Internet censorship, which has also been studied, particularly with regard to the Great Firewall of China (GFC). The ConceptDoppler project [8] provides a "censorship weather report" by detecting the synthetic TCP reset packets, tracking which words are blocked when they are detected in TCP streams. Park and Crandall [25] investigated the distributed nature of the filtering of HTTP responses in China. They showed a low disconnection successful rate (<51%) when the censor detected a keyword and attempted to reset the connection. Clayton et al. [6, 7], studied the hybrid blocking system deployed by British Telecom in the United Kingdom to block access to pedophile websites and also investigated the GFC. Wolfgarten [35] studied filtering situation in China and also showed some bypass techniques. Wang [34], compared the search return results from China with the results from New Zealand, similar to our own methodology, but with only 200 English keywords. And, of course, a wide variety of technical countermeasures to the GFC and other sources of censorship have been designed and deployed, most notably Tor [9]. Likewise, a wide variety of peer-to-peer networks have been proposed with censorship-resistant and/or anonymization properties (see, e.g., [30, 12, 22, 23, 13, 27]).

Besides computer scientists, Internet censorship also attract attention from other fields. MacKinnon [21] studied how Chinese blog services censor their bloggers. Zittrain and Edelman [38] investigated Internet filtering methods and collected lists of sites blocked by China. Other studies have been conducted by a variety of public interest organizations including Global Internet Freedom Consortium [14], Human Rights Watch [17], Villenevue [33], and Reports Without Borders [31].

# 7. CONCLUSIONS AND FUTURE WORK

From our experiments and analysis, we can now conclude:

- The Chinese government does not necessarily make unambiguous censorship requirements of search engine companies, although we have seen evidence of a "black list" that they must follow.
- Search engine companies appear to decide, on their own, how they might perform censorship. The terms which are filtered and the precise methodology with which they are filtered clearly vary across the different search engines, but there are clear topic areas that are always subject to censorship: political activists and their organizations, particularly anything to do with the Tiananmen Square protests or Falun Gong.
- Some search engines maintain "white lists" of web sites that are considered "safe" for responses to any blacklisted search query.
- Chinese search engines also aggressively filter pornographic terms, although their strategies toward doing this vary and are not necessarily very sophisticated.
- Non-Chinese-language search engines also implement anti-pornography filtering, but not political censorship when given queries in Chinese languages.
- The Great Firewall of China mainly blocks connections when it sees political sensitive terms. Rather than a once-size-fits-all policy, different services within China are subject to differing levels of blocking.
- The methodologies introduced in this paper may help users to recognize a term being censored. Users can exploit a variety of telltale signs to, at the very least, know with certainty that they are looking at unnaturally filtered search engine results.
- Censored search terms are often treated differently when queried within double quotation marks. This may be a partial mitigation against some search engine censorship.

Overall, we demonstrated a variety of straightforward techniques and measurements that can be used to quantify Chinese Internet censorship. Of course, if China wished, it would be straightforward for its search engines to hide many of the features that we used. Hit counts could be falsified. Censorship within a given query need not be reported. Unfortunately, this makes our results too fragile to be of much use to Chinese citizens wishing to overcome search engine censorship.

Regardless, we see a variety of interesting challenges for future work. For example, we could conduct a far more invasive comparison of search results between, for example, Google.com and Baidu.com, looking at the size of the set intersection of their top 1000 results or other such comparison metrics. Such an analysis, conducted at higher frequency and for a deeper list of queries, would require a much larger network of computers to capture the data and analyze the results. Among other things, we would certainly need a larger pool of IP addresses to avoid search engines' query rate limiting and to overcome the inevitable IP blacklisting we would experience.

With a deeper and more frequent set of queries, we could have perhaps been able to have detailed traces leading up to an important political event, such as the recent detainment of Chinese artist and political dissident Ai Weiwei. Likewise, we might have been able to observe the presence or absence of censorship of news from other countries where the governments are under duress (e.g., Libya and Syria) or have been overthrown (e.g., Tunisia and Egypt). If censorship policies turn out to be adjusted in advance of other government actions, that would be a very valuable signal.

Of course, being able to measure Chinese Internet censorship from the outside, or to measure the censorship of many other countries with comparably restricted Internet access, is of only limited utility to the citizens inside those countries. Still, we did find one time when the Great Firewall briefly dropped its guard. If we could effectively and efficiently identify these opportunities, for the limited durations when they may exist, this could lead to pointed action, directed toward arranging external content to most effectively communicate to those on the inside, reaching out to learn more.